\documentclass[draftcls,11pt,onecolumn,romanappendices]{IEEEtran}
\usepackage{mathpple}
\usepackage{times}
\usepackage{amsthm,xpatch}
\usepackage{amsmath,amsfonts}
\usepackage{cite}
\usepackage{amssymb}
\usepackage{dsfont}
\usepackage{graphicx, subfigure}
\usepackage{color}
\usepackage{breqn}
\usepackage{mathtools}
\usepackage{bbm}
\usepackage{latexsym}
\usepackage[ruled, linesnumbered]{algorithm2e}
\usepackage{accents}
\usepackage{tikz}
\usepackage{scalerel}
\usepackage{amstext,amsthm,amsmath,amsfonts,latexsym,graphicx,amssymb,epsfig,epsf,psfrag,mathtools}
\usepackage{pgfpages,xcolor,subfigure,pstricks,color}
\usepackage{pgfplots}
\usepackage{multirow}

\DeclareMathOperator*{\argmax}{\arg\!\max}

\newcommand*{\transpose}{%
  {\mathpalette\@transpose{}}%
}

\IEEEoverridecommandlockouts

\begin{document}

\newcommand{\SB}[3]{
\sum_{#2 \in #1}\biggl|\overline{X}_{#2}\biggr| #3
\biggl|\bigcap_{#2 \notin #1}\overline{X}_{#2}\biggr|
}

\newcommand{\Mod}[1]{\ (\textup{mod}\ #1)}

\newcommand{\overbar}[1]{\mkern 0mu\overline{\mkern-0mu#1\mkern-8.5mu}\mkern 6mu}

\makeatletter
\newcommand*\nss[3]{%
  \begingroup
  \setbox0\hbox{$\m@th\scriptstyle\cramped{#2}$}%
  \setbox2\hbox{$\m@th\scriptstyle#3$}%
  \dimen@=\fontdimen8\textfont3
  \multiply\dimen@ by 4             
  \advance \dimen@ by \ht0
  \advance \dimen@ by -\fontdimen17\textfont2
  \@tempdima=\fontdimen5\textfont2  
  \multiply\@tempdima by 4
  \divide  \@tempdima by 5          
  \ifdim\dimen@<\@tempdima
    \ht0=0pt                        
    \@tempdima=\fontdimen5\textfont2
    \divide\@tempdima by 4          
    \advance \dimen@ by -\@tempdima 
    \ifdim\dimen@>0pt
      \@tempdima=\dp2
      \advance\@tempdima by \dimen@
      \dp2=\@tempdima
    \fi
  \fi
  #1_{\box0}^{\box2}%
  \endgroup
  }
\makeatother

\makeatletter
\renewenvironment{proof}[1][\proofname]{\par
  \pushQED{\qed}%
  \normalfont \topsep6\p@\@plus6\p@\relax
  \trivlist
  \item[\hskip\labelsep
        \itshape
    #1\@addpunct{:}]\ignorespaces
}{%
  \popQED\endtrivlist\@endpefalse
}
\makeatother

\makeatletter
\newsavebox\myboxA
\newsavebox\myboxB
\newlength\mylenA

\newcommand*\xoverline[2][0.75]{%
    \sbox{\myboxA}{$\m@th#2$}%
    \setbox\myboxB\null
    \ht\myboxB=\ht\myboxA%
    \dp\myboxB=\dp\myboxA%
    \wd\myboxB=#1\wd\myboxA
    \sbox\myboxB{$\m@th\overline{\copy\myboxB}$}
    \setlength\mylenA{\the\wd\myboxA}
    \addtolength\mylenA{-\the\wd\myboxB}%
    \ifdim\wd\myboxB<\wd\myboxA%
       \rlap{\hskip 0.5\mylenA\usebox\myboxB}{\usebox\myboxA}%
    \else
        \hskip -0.5\mylenA\rlap{\usebox\myboxA}{\hskip 0.5\mylenA\usebox\myboxB}%
    \fi}
\makeatother

\xpatchcmd{\proof}{\hskip\labelsep}{\hskip3.75\labelsep}{}{}

\pagestyle{plain}

\title{\fontsize{21}{28}\selectfont 
Two-Stage Adaptive Pooling with RT-qPCR for COVID-19 Screening
}

\author{Anoosheh Heidarzadeh and Krishna Narayanan\\
Department of Electrical and Computer Engineering \\
Texas A\&M University\\
College Station, TX 77843\thanks{This material is based upon work supported by the National Science Foundation (NSF) under Grant CCF-2027997.}}

\maketitle 

\thispagestyle{plain}



\begin{abstract}
We propose two-stage adaptive pooling schemes, 2-STAP and 2-STAMP, for detecting COVID-19 using real-time reverse transcription quantitative polymerase chain reaction (RT-qPCR) test kits.
Similar to the Tapestry scheme of Ghosh \emph{et al.}, the proposed schemes leverage soft information from the 
RT-qPCR process about the total viral load in the pool. 
This is in contrast to conventional group testing schemes where the measurements are Boolean.
The proposed schemes provide higher testing throughput than the popularly used Dorfman's scheme. 
They also provide higher testing throughput, sensitivity and specificity than the 
state-of-the-art non-adaptive Tapestry scheme.
The number of pipetting operations is lower than state-of-the-art non-adaptive pooling schemes, and is higher than that for the Dorfman's scheme. 
The proposed schemes can work with substantially smaller group sizes than non-adaptive schemes and are simple to describe. 
Monte-Carlo simulations using the statistical model in the work of Ghosh \emph{et al.} (Tapestry) show that 10 infected people in a population of size 961 can be identified with 70.86 tests on the average with a sensitivity of 99.50\% and specificity of 99.62\%. 
This is 13.5x, 4.24x, and 1.3x the testing throughput of individual testing, Dorfman's testing, and the Tapestry scheme, respectively.
\end{abstract}


\section{Introduction}
There is broad consensus among epidemiologists, economists and policy makers that wide-scale testing of asymptomatic patients is the key for reopening the economy. While the benefits of testing are obvious, shortage of testing kits, reagents and the ensuing low-throughput of individual testing protocols has prevented deployment of wide-scale testing.
Group testing or, pooling is an alternative way to substantially increase the testing throughput. 

The idea of group testing was introduced by Dorfman \cite{dorfman1943detection} during World War II for testing soldiers for syphilis without having to test each soldier individually. 
Dorfman's scheme consists of two stages (or rounds). In the first stage, the set of people to be tested is split into disjoint pools and a test is performed on each pool. If a pool tested negative, everyone in that pool will be identified as non-infected. Otherwise, if a pool tested positive, we proceed to the second stage where all people in a positive pool will be tested individually, and then identified as infected or non-infected accordingly. 
When the prevalence is small, Dorfman's scheme requires substantially fewer tests than individual testing.

Dorfman-style testing has been implemented in the past in screening for many diseases including HIV~\cite{emmanuel1988pooling}, Chlamydia and Gonorrhea~\cite{lewis2012cost}. It has also been considered for screening for influenza \cite{van2012pooling}. 
For COVID-19, several experimental results have confirmed the feasibility of using Dorfman-style pooling and it has been implemented in Nebraska, Germany, India and China \cite{bilder2020tests}, \cite{shental2020efficient}, \cite{Yelin2020}, \cite{hogan2020sample}.

While Dorfman-style pooling is easy to implement, it is not optimal. 
Over the past 75 years, more sophisticated group testing schemes that provide higher testing throughput have been designed.
The literature on group testing is too vast to review in detail and an overview of the techniques can be found in \cite{du2000combinatorial} and \cite{AJS2019}. 
Group testing is also related to compressed sensing and insights from compressed sensing have been used to design group testing schemes. 
An important difference between group testing and compressed sensing is that in group testing, 
the measurements are Boolean (test result is either positive or negative) and 
they naturally correspond to \emph{non-linear} functions of the unknown vector. 

The vast majority of the work using group testing with real-time reverse transcription quantitative polymerase chain reaction (RT-qPCR) has only considered Boolean measurements even though the RT-qPCR process can produce more fine-grained information (soft information) about the total viral load in the pool. 
It is well-known in information theory that such soft information can potentially be used to increase testing throughput substantially. 
However, group testing schemes that leverage soft information from the RT-qPCR process remain largely unexplored.

Very recently, Ghosh {\em et al.} in \cite{ghosh2020tapestry} developed a statistical model relating the soft information from the RT-qPCR to the total viral load in the pool.
They designed a scheme called Tapestry, which uses \emph{non-adaptive} group testing using Kirkman triples
and they considered several decoding algorithms that use the soft information. 
They showed substantial gains in testing throughput over Dorfman's scheme and to the best of our knowledge, this scheme is the state of the art non-adaptive group testing scheme that works with RT-qPCR, especially since it is the only work we are aware of that uses the soft information from the RT-qPCR measurement process. 

Here, we propose two simple and effective \emph{two-stage adaptive pooling} schemes that use the soft information from the RT-qPCR process
and provide several advantages over Dorfman's scheme and the Tapestry scheme.
We refer to these algorithms as the Two-stage Adaptive Pooling (2-STAP) and the Two-stage Adaptive Mixed Pooling (2-STAMP) schemes/algorithms.
The proposed schemes provide substantially higher throughput than Dorfman-style testing.
Compared to the Tapestry scheme in \cite{ghosh2020tapestry}, 2-STAP and 2-STAMP have higher testing throughput and under the statistical model developed in \cite{ghosh2020tapestry}, for all tested cases, our algorithms have higher sensitivity and higher specificity.
The proposed algorithms require fewer pipetting operations than Tapestry, but require more pipetting operations than Dorfman's scheme.
Finally, 2-STAP and 2-STAMP work with much smaller pool sizes and population sizes than the Tapestry algorithm and hence, is easy to describe and
implement in the lab. 
Monte-Carlo simulations using the statistical model in the work of Ghosh \emph{et al.} (Tapestry), show that 10
infected people in a population of size 961 can be identified with 70.86 tests on the average with a sensitivity of 99.50\% and specificity of 99.62\% with a pool size of 31. This is 13.5x, 4.24x, and 1.3x the testing throughput of individual testing, Dorfman's testing, and the Tapestry scheme, respectively. 

Unlike Tapestry, which is a non-adaptive scheme, 2-STAP and 2-STAMP require storage of the swab samples and their accessibility for the second round of testing---similar to that of Dorfman's scheme.


\section{Problem Setup}
In this section, we explain the problem setup. Throughout, we will consider the example of pooling-based testing for COVID-19 using the real-time reverse transcription quantitative polymerase chain reaction (RT-qPCR) technique---considered also in~\cite{ghosh2020tapestry}---as an application of sensing with binary matrices for support recovery of sparse signals. 

Let $\mathbb{R}_{\geq 0} = \{x\in \mathbb{R}: x\geq 0\}$ and $\mathbb{R}_{>0} = \mathbb{R}_{\geq 0}\setminus \{0\}$. 
For any integer $i> 0$, we denote $\{1,\dots,i\}$ by $[i]$, and define $[0] = \emptyset$. 

Consider a population of $n$ people, labeled $1,\dots,n$, that are to be tested for COVID-19. The vector of viral loads of these people can be modeled by a signal $\mathbf{x} = [x_1,\dots,x_n]^{\mathsf{T}}$, $x_j\in \mathbb{R}_{\geq 0}$, where the $j$th coordinate of $\mathbf{x}$ represents the viral load of the $j$th person. If the $j$th person is infected (i.e., COVID-19 positive), then $x_j$ is a nonzero value; otherwise, if the $j$th person is not infected (i.e., COVID-19 negative), then $x_j$ is zero. 
We assume that every coordinate in $\mathbf{x}$ is nonzero with probability $p$ (or zero with probability ${1-p}$), independently from other coordinates, and 
every nonzero coordinate takes a value from $\mathbb{R}_{>0}$ according to a fixed and known probability distribution $p_x$. 
Note that the sparsity parameter $p$ may or may not be known. In this context, the sparsity parameter $p$ is known as \emph{prevalence}.


We denote the number of nonzero coordinates in $\mathbf{x}$ (e.g., the number of infected people) by $n_{+}$, and denote the number of zero coordinates in $\mathbf{x}$ (e.g., the number of non-infected people) by $n_{-}$. 
Note that $n_{+}+n_{-} = n$. 
We denote by $S(\mathbf{x})$ the support set of $\mathbf{x}$, i.e., the index set of all nonzero coordinates in $\mathbf{x}$. Note that $|S(\mathbf{x})| = n_{+}$.

The $i$th binary linear measurement $y$ of $\mathbf{x}$ is defined as a linear combination of coordinates $x_j$'s according to the coefficients $a_{ij}$'s that are elements from $\{0,1\}$. 
That is, $y_i = \mathbf{a}_i\cdot \mathbf{x} =  \sum_{j=1}^{n} a_{i,j} x_j$, where $\mathbf{a}_i = [a_{i,1},\dots,a_{i,n}]$, $a_{i,j} \in \{0,1\}$. (Note that a binary linear measurement is different from a Boolean measurement. In the former, the coefficients are binary but the measurement value can be a real number, whereas in the latter both the coefficients and the measurement value are binary.) 
For example, a measurement $y_i$ represents the sum of viral loads of a subset of people to be tested for COVID-19.
Given a measurement $y_i$, any coordinate $x_{ij}$ such that $a_{i,j}=1$ (or $a_{i,j}=0$) is referred to as an \emph{active (or inactive) coordinate} in the measurement $y$. 

Suppose we sense the signal $\mathbf{x}$ by making the measurements $y_1,y_2,\dots$, and observe noisy versions of $y_1,y_2,\dots$, denoted by $z_1,z_2,\dots$. 
The $i$th measurement $y_i$ and the noisy measurement $z_i$ are given by 
\begin{eqnarray}
\nonumber
    y_i & = & \sum_{j=1}^{n} a_{i,j} x_j\\
\label{eqn:multiplicativenoise}
    z_i & = & y_i\varepsilon_i,
\end{eqnarray}
where $\varepsilon_i$'s are independent realizations of a random variable $\varepsilon$ -- taking values from $\mathbb{R}_{>0}$ according to a fixed and known probability distribution $p_{\varepsilon}$. 
(The reason we consider a multiplicative noise model, instead of the commonly-used additive noise model, will be discussed shortly.)  
Note that $z_i = 0$ if and only if $y_i = 0$ (i.e., all active coordinates in the $i$th measurement are zero coordinates), and $z_i\neq 0$ if and only if $y_i\neq 0$ (i.e., there exists at least one nonzero coordinate among the active coordinates in the $i$th measurement). A detailed explanation about the multiplicative noise model in~\eqref{eqn:multiplicativenoise} can be found in Appendix~\ref{app:0}.

Our goal is to collect as few noisy measurements $z_1,z_2,\dots$ as possible for any signal $\mathbf{x}$ such that the support set $S(\mathbf{x})$ can be recovered from $z_1,z_2,\dots$, with a target level of accuracy as defined shortly. 

We refer to the process of generating the measurements as \emph{sensing}, and refer to the process of estimating the support set from the noisy measurements as \emph{(signal-support) recovery}. 
Given a sensing algorithm and a recovery algorithm, we denote the estimate of $S(\mathbf{x})$ by $\widehat{S}(\mathbf{x})$, which depends on the noisy measurements $z_1,z_2,\dots$ and the sensing and recovery algorithms. Any coordinate $x_j$ such that $j\in S\setminus \widehat{S}(\mathbf{x})$ is referred to as a \emph{false negative}, and any coordinate $x_j$ such that $j\in \widehat{S}(\mathbf{x})\setminus S(\mathbf{x})$ is referred to as a \emph{false positive}. 
Similarly, any coordinate $x_j$ such that $j\notin S(\mathbf{x})\cup \widehat{S}(\mathbf{x})$ is referred to as a \emph{true negative}, and any coordinate $x_j$ such that $j\in S(\mathbf{x})\cap \widehat{S}(\mathbf{x})$ is referred to as a \emph{true positive}. 
We denote by $f_{-}(\mathbf{x})$ the number of false negatives, i.e., $f_{-}(\mathbf{x}) = |S(\mathbf{x})\setminus \widehat{S}(\mathbf{x})|$. 
Similarly, we denote by $f_{+}(\mathbf{x})$ the number of false positives, i.e., $f_{+}(\mathbf{x}) = |\widehat{S}(\mathbf{x})\setminus S(\mathbf{x})|$. 
Note that $f_{-}(\mathbf{x})$ and $f_{+}(\mathbf{x})$ depend on the noisy measurements $z_1,z_2,\dots$ and the sensing and recovery algorithms. 

Given a sensing algorithm and a recovery algorithm, the \emph{false negative rate} $r_{-}$ is defined as the expected value of the ratio of the number of false negatives to the number of nonzero coordinates, i.e., $r_{-} = \mathbb{E}[f_{-}(\mathbf{x})/n_{+}(\mathbf{x})]$, where the expectation is taken over all signals $\mathbf{x}$. 
Similarly, the \emph{false positive rate} $r_{+} = \mathbb{E}[f_{+}(\mathbf{x})/n_{-}(\mathbf{x})]$. 
It should be noted that the ratios $f_{-}(\mathbf{x})/n_{+}(\mathbf{x})$ and $f_{+}(\mathbf{x})/n_{-}(\mathbf{x})$ are random variables, because they depend on $\mathbf{x}$, which is itself random in both the deterministic and probabilistic models defined earlier. 
Also, conditioned on $\mathbf{x}$ having $k$ nonzero coordinates and $n-k$ zero coordinates, we denote the \emph{conditional false negative rate} by ${r}_{-}^{[k]}$ and the \emph{conditional false positive rate} by ${r}_{+}^{[k]}$. 
The quantities $1-r_{-}$ and $1-r_{+}$ are known as (unconditional) \emph{sensitivity} and \emph{specificity}, respectively. Analogously, we refer to $1-r^{[k]}_{-}$ and $1-r^{[k]}_{+}$ as \emph{conditional sensitivity} and \emph{conditional specificity}, respectively. 

For given thresholds ${0\leq \delta_{-},\delta_{+}<1}$, our goal is to design a sensing algorithm and a recovery algorithm such that with minimum number of measurements the constraints ${r}_{-}\leq \delta_{-}$ and ${r}_{+}\leq \delta_{+}$ (or ${r}^{[k]}_{-}\leq \delta_{-}$ and ${r}^{[k]}_{+}\leq \delta_{+}$) are satisfied. 
The thresholds $\delta_{-}$ and $\delta_{+}$ specify the target level of accuracy for support recovery.  

\section{Single-Stage Schemes versus Multi-Stage Schemes}
In a single-stage sensing scheme, also known as \emph{non-adaptive sensing}, $m$ measurements $y_1,\dots,y_m$ are made in parallel, and $m$ noisy measurements $z_1,\dots,z_m$ are observed. 
The coefficient vectors of the measurements $y_1,\dots,y_m$ can be represented by an $m\times n$ sensing matrix $\mathbf{A}$ with entries from $\{0,1\}$. 
That is, $\mathbf{A} = [\mathbf{a}_1^{\mathsf{T}},\dots,\mathbf{a}_m^{\mathsf{T}}]^{\mathsf{T}}$ where $\mathbf{a}_i = [a_{i,1},\dots,a_{i,n}]$ represents the $i$th row of the sensing matrix $\mathbf{A}$, i.e., the coefficients of $x_j$'s in the $i$th measurement $y_i$. 
It should be noted that in a single-stage sensing scheme, the sensing matrix $\mathbf{A}$ is designed in advance, prior to any initial sensing of the signal, and hence the name ``non-adaptive sensing''. We denote the vector of measurements by $\mathbf{y} = [y_1,\dots,y_m]^{\mathsf{T}}$, the vector of noisy measurements by $\mathbf{z} = [z_1,\dots,z_m]^{\mathsf{T}}$, and the vector of noise values by $\boldsymbol{\varepsilon} = [\varepsilon_1,\dots,\varepsilon_m]^{\mathsf{T}}$. 
Then, $\mathbf{y} = \mathbf{A}\mathbf{x}$, and $\mathbf{z} = \mathbf{y}\circ \boldsymbol{\varepsilon}$, where the symbol ``$\circ$'' denotes the element-wise (Hadamard) product. 
Given a sensing matrix $\mathbf{A}$, a recovery algorithm seeks to compute an estimate $\widehat{S}(\mathbf{x})$ of the support set $S(\mathbf{x})$ of the signal $\mathbf{x}$ from the noisy measurement vector $\mathbf{z}$. 

The idea of single-stage sensing can be extended to $T$-stage sensing (for any natural number $T$) as follows. 
A $T$-stage sensing scheme consists of $T$ sensing matrices $\mathbf{A}^{(1)},\dots,\mathbf{A}^{(T)}$, where $\mathbf{A}^{(t)}$ is an $m^{(t)}\times n$ matrix with entries from $\{0,1\}$. 
Similarly as in the case of single-stage sensing scheme, for each stage $t$ we denote the vector of measurements by $\mathbf{y}^{(t)} = [y^{(t)}_1,\dots,y^{(t)}_{m^{(t)}}]^{\mathsf{T}}$, the vector of noisy measurements by $\mathbf{z}^{(t)} = [z^{(t)}_1,\dots,z^{(t)}_{m^{(t)}}]^{\mathsf{T}}$, and the vector of noise values by $\boldsymbol{\varepsilon}^{(t)} = [\varepsilon^{(t)}_1,\dots,\varepsilon^{(t)}_{m^{(t)}}]^{\mathsf{T}}$. 
Then, $\mathbf{y}^{(t)} = \mathbf{A}^{(t)}\mathbf{x}$ and $\mathbf{z}^{(t)} = \mathbf{y}^{(t)}\circ \boldsymbol{\varepsilon}^{(t)}$. 
For each $t>1$, the measurements in the $t$th stage are all made in parallel, similar to the single-stage case, but the design of sensing matrix $\mathbf{A}^{(t)}$ depends on the design of all sensing matrices $\mathbf{A}^{(1)},\dots,\mathbf{A}^{(t-1)}$ and all noisy measurement vectors $\mathbf{z}^{(1)},\dots,\mathbf{z}^{(t-1)}$. Given the sensing matrices $\mathbf{A}^{(1)},\dots,\mathbf{A}^{(T)}$, a recovery algorithm seeks to compute an estimate $\widehat{S}(\mathbf{x})$ of $S(\mathbf{x})$ from $\mathbf{z}^{(1)},\dots,\mathbf{z}^{(T)}$.     

\section{Related Problems}
The similarities and differences between this work and those in the literature of Compressed Sensing (CS) and Quantitative Group Testing (QGT) are summarized below: 
\begin{itemize}
    \item \emph{Binary versus non-binary signals:} 
    In QGT, the nonzero coordinates of the signal have all the same value; 
    whereas in CS -- similar to the present work, the nonzero coordinates can take different values.   
    
    \item \emph{Real versus binary sensing matrices:} 
    Many of the existing CS algorithms rely on sensing matrices with real entries; 
    whereas the sensing matrices in the QGT algorithms -- similar to this work, are binary. 
    
    \item \emph{Unconstrained versus constrained sensing matrices:} 
    In CS and QGT, there is often no explicit constraint on the number of nonzero entries per row (referred to as the row weight) and the number of nonzero entries per column (referred to as the column weight) of the sensing matrices. 
    However, in this work, the focus is on sensing matrices in which the row weights and the column weights are constrained.    
    
    \item \emph{Additive versus multiplicative noise:} 
    Almost all existing work on CS and QGT consider either the noiseless setting, or the settings in which the noise is additive. 
    The focus of this work is, however, on the multiplicative noise. 
    
    \item \emph{Single-stage versus multi-stage sensing:} 
    Most of the existing work on CS focus on single-stage (non-adaptive) sensing algorithms. 
    However, in QGT -- similar to this work, both single-stage and multi-stage sensing algorithms have been studied. 
    
    \item \emph{Signal recovery versus signal-support recovery:} 
    In QGT, the signal recovery and the support recovery problems are equivalent, since the signal is binary-valued. 
    In CS, however, the goal is often to estimate the signal itself, rather than estimating the support of the signal only. 
    In this work, the goal is to recover the support of the signal (similar to QGT), even though the signal is non-binary (similar to CS). 
\end{itemize}

\section{Related Work: Tapestry}
For the same setting as the one in this work, a scheme called \emph{Tapestry} was recently proposed in~\cite{ghosh2020tapestry}. 
In what follows, we briefly explain the sensing and recovery algorithms in this scheme. 

\subsection{Sensing Algorithms}
The Tapestry scheme employs a single-stage sensing algorithm. 
The sensing matrices used in this scheme are based on the Kirkman triples, a special class of the Steiner triples, borrowed from the combinatorial design theory. 

Let $m$ be a multiple of $3$, and let $c$ be an integer such that $c\leq \frac{m-1}{2}$. 
A (balanced) Kirkman triple system (with parameters $m$ and $c$) can be represented by a binary matrix with $m$ rows and $\frac{m}{3}\times c$ columns such that: 
\begin{itemize}
    \item The support set of each column has size exactly $3$; 
    \item The intersection of support sets of any two columns has size at most $1$;
    \item For any $0\leq i\leq \frac{m}{3}(c-1)$ such that $i$ is a multiple of $\frac{m}{3}$, the sum of the columns indexed by $i+1, i+2,\dots,i+\frac{m}{3}$ is equal to an all-one vector.
\end{itemize}

\subsection{Recovery Algorithm}
In~\cite{ghosh2020tapestry}, several different recovery algorithms from the CS literature were considered. 
In the following we explain one of these recovery algorithms which will also be used as part of the proposed schemes in this paper, and refer the reader to Appendix~\ref{app:I} for detailed description of the other recovery algorithms considered in~\cite{ghosh2020tapestry}.      
\subsubsection*{Combinatorial Orthogonal Matching Pursuit (COMP)}
The COMP algorithm -- which will also be used as part of the proposed schemes in this paper, works based on a simple observation: 
if $z_i = 0$ (and hence, $y_i=0$ since the noise $\varepsilon_i$ takes only nonzero values), then all active coordinates in the $i$th measurement are zero coordinates. 
The algorithm initially marks all coordinates $x_j$'s as ``unknown''. Then, the algorithm finds all $i$ such that $z_i = 0$. Let $I = \{i\in [m]: z_i=0\}$. For every $i\in I$, the algorithm marks all active coordinates in the $i$th measurement as ``known''. Upon termination of the algorithm, the index set of all coordinates that are marked ``unknown'' yields an estimate $\widehat{S}(\mathbf{x})$ of the support set $S(\mathbf{x})$. 
(Any coordinate marked ``known'' is surely a zero coordinate in the signal, and its index will not be included in the estimated support set.) 
Note that the support set estimated by the COMP algorithm contains no false negatives, but it may contain some false positives. That is, $S(\mathbf{x})\subseteq \widehat{S}(\mathbf{x})$, or in other words,  $\widehat{S}(\mathbf{x})$ is a superset estimate of $S(\mathbf{x})$.

\section{2-STAP: A Two-Stage Adaptive Pooling}\label{subsec:TSSwithoutCP}
In this section, we propose a two-stage sensing algorithm and an associated recovery algorithm which we collectively refer to as the 2-STAP scheme. 
The proposed scheme is inspired by the well-known two-stage Dorfman's scheme which was originally proposed in the context of group testing but requires substantially fewer measurements. 
Translating the Dorfman's scheme into the language of our work, in the first stage, the signal coordinates are pooled into a number of disjoint groups of equal size, and one measurement is made for each pool where all coordinates in the pool are active in the measurement. 
In the second stage of Dorfman's scheme, one measurement is made for every coordinate in a \emph{positive pool} (i.e., a pool whose measurement in the first stage is nonzero), and no additional measurements are made for any \emph{negative pool} (i.e., a pool whose measurement in the first stage is zero).

The first stage of the 2-STAP scheme is the same as that of Dorfman's scheme. 
The second stage, however, differs from that of Dorfman's scheme. 
In particular, unlike the second stage of Dorfamn's scheme, in the second stage of the 2-STAP scheme, a number of measurements are made for different (not necessarily singleton) subsets of coordinates, in each positive pool. 
Similar to the Dorfman's scheme, the second stage of the 2-STAP scheme makes no additional measurements for any negative pools. 


In the following, we denote by $\mathbf{1}_{t}$ or $\mathbf{0}_{t}$ an all-one or an all-zero row vector of length $t$, respectively. 


Given a signal $\mathbf{x}$, we partition the $n$ signal coordinates $x_1,\dots,x_n$ into $q$ pools of size $s = n/q$. 
We denote by $\mathbf{x}_{l}$ the $l$th pool of coordinates, i.e., 
$\mathbf{x}_l = [x_{(l-1)s+1},\dots,x_{ls}]^{\mathsf{T}}$. 

We denote by $m^{(1)}_l$ and $m^{(2)}_l$ the number of measurements for the $l$th pool in the first and second stage, respectively. 
Also, we denote by $\mathbf{A}^{(1)}_l$ and $\mathbf{A}^{(2)}_l$ the sensing matrix corresponding to the $l$th pool in the first and second stage, respectively, and 
denote by $\mathbf{A}_l = [(\mathbf{A}^{(1)}_l)^{\mathsf{T}},(\mathbf{A}^{(2)}_l)^{\mathsf{T}}]^{\mathsf{T}}$ the overall sensing matrix corresponding to the $l$th pool. 
Note that $\mathbf{A}^{(1)}_l$ is an $m^{(1)}_l\times s$ matrix, $\mathbf{A}^{(2)}_l$ is an $m^{(2)}_l\times s$ matrix, and $\mathbf{A}_l$ is an $m_l\times s$ matrix, where $m_l = m^{(1)}_l+m^{(2)}_l$.  
Let $m^{(1)} = \sum_{l\in [q]} m^{(1)}_l$ and $m^{(2)} = \sum_{l\in [q]} m^{(2)}_l$. 
Note that $m^{(1)}$ and $m^{(2)}$ are the total number of measurements in the first and second stage, respectively.   

\subsection{Sensing Algorithm for First Stage}
In the first stage, for each pool $l\in [q]$, we make one measurement $y^{(1)}_l = \mathbf{1}_{s}\cdot \mathbf{x}_l$. 
That is, $m^{(1)}_l = 1$ and $\mathbf{A}^{(1)}_l = \mathbf{1}_{s}$. 
Thus, the total number of measurements in the first stage is $m^{(1)}=q$, and the sensing matrix in the first stage, $\mathbf{A}^{(1)}$, is a $q\times n$ matrix whose $l$th row is $\mathbf{a}^{(1)}_l = [\mathbf{0}_{(l-1)s},\mathbf{1}_{s},\mathbf{0}_{(q-l)s}]$. 

\subsection{Recovery Algorithm for First Stage}
Suppose we observe the noisy measurements $z^{(1)}_l = y^{(1)}_l\varepsilon^{(1)}_l$ for $l\in [q]$. 
Let $L = \{l\in [q]: z^{(1)}_l= 0\}$. 
Assume, without loss of generality (w.l.o.g.), that $L = [q]\setminus [t]$ for some $0\leq t\leq q$. 
That is, the first $t$ pools are all positive, and the last $q-t$ pools are all negative. 

\subsection{Sensing Algorithm for Second Stage}
For any $l\in [t]$, we need to make additional measurements for the $l$th pool in the second stage, i.e., $m^{(2)}_l>0$ for all $l\in [t]$, because such a pool contains at least one nonzero coordinate.
For any $l\in [q]\setminus [t]$, the $l$th pool contains only zero coordinates, and we do not need to make any additional measurements for any such pool in the second stage, i.e., $m^{(2)}_l=0$ for all $l\in [q]\setminus [t]$. Below, we focus on the positive pools, i.e., the pools $1,\dots,t$. 

Intuitively, the larger is $m^{(2)}_l$, the smaller will be the (conditional) false negative/positive rate, but the larger will be the average number of measurements. Since it is not known how to theoretically optimize $m^{(2)}_l$, we resort to a heuristic approach to choose $m^{(2)}_l$. 
In particular, we present two variants of the 2-STAP scheme: 2-STAP-I and 2-STAP-II.
In 2-STAP-I, for all positive pools, the number of measurements and the pooling scheme in the second stage will be the same, regardless of the observed measurements for these pools in the first stage. 
That is, $m^{(2)}_l$ is fixed for all $l\in [t]$. 
On the other hand, in 2-STAP-II, for each positive pool, the number of measurements and the pooling scheme in the second stage will be chosen based on 
the number of nonzero coordinates in $\mathbf{x}_l$, denoted by $k_l$.  Note that $k_l$ may not be known \emph{a priori}, but an estimate $\widehat{k}_l$ of $k_l$ can be computed using the observed measurement for the $l$th pools in the first stage, as follows. 


Let $p(k)$ be the probability that $\mathbf{x}_l$ has $k$ nonzero coordinates and $s-k$ zero coordinates, and 
let $p(z^{(1)}_l|k)$ be the probability density of $z_l^{(1)}=y_l^{(1)}\varepsilon^{(1)}_l$ where $y^{(1)}_l=\mathbf{1}_s\cdot \mathbf{x}_l$ given that $\mathbf{x}_l$ has $k$ nonzero coordinates and $s-k$ zero coordinates. 
If the sparsity parameter $p$ is known, for any $k$, $p(k)$ and $p(z^{(1)}_l|k)$ can be computed exactly or approximately (depending on the distribution of values of the nonzero coordinates in the signal and the distribution of the noise). 
In this case, given a noisy measurement $z^{(1)}_l$, a maximum-a-posteriori (MAP) estimate of $k_l$ is given by $\widehat{k}_l = \argmax_{k} p(k) p(z^{(1)}_l|k)$. 
If $p$ is not known, we first compute a maximum-likelihood (ML) estimate $\widehat{p} = 1-(1-\frac{t}{q})^{\frac{1}{s}}$ of $p$, and 
then use $\widehat{p}$, instead of $p$, to first compute $p(k)$ and $p(z^{(1)}_l|k)$ for any $k$, and then compute a MAP estimate $\widehat{k}_l$ of $k_l$. 
(Note that the average number of positive pools is $q(1-(1-p)^s)$, and $t$ is a realization of the number of positive pools. 
Setting these quantities equal to each other and solving for $p$, we get the estimate $\widehat{p}$ of $p$, as defined above.)   

Given $m^{(2)}_l$, the optimal design of $\mathbf{A}^{(2)}_l$ is not known. In this work, for each $l\in [t]$, we randomly choose $\mathbf{A}^{(2)}_l$ from the ensemble of all $m^{(2)}_l\times s$ binary matrices (with distinct rows and distinct columns) with a pre-specified row/column weight profile. The weight profile should be chosen to obtain a good trade-off between the computational complexity (of the sensing and recovery algorithms) and false negative/positive rates. 
The weight profile of matrices used in our simulations can be found in Appendix~\ref{app:I}.


\subsection{Recovery Algorithm for Second Stage}
The recovery algorithm in the second stage consists of three steps: COMP algorithm, MAP decoding, and list generation. 

For each $l\in [t]$, suppose the noisy measurement vector $\mathbf{z}^{(2)}_l = \mathbf{y}^{(2)}_l\boldsymbol{\varepsilon}^{(2)}_l$ is observed, where $\mathbf{y}^{(2)}_l = \mathbf{A}^{(2)}_l\mathbf{x}_l$. 
Let $\mathbf{z}_l = [z^{(1)}_l,(\mathbf{z}^{(2)}_l)^{\mathsf{T}}]^{\mathsf{T}}$ be the overall noisy measurement vector corresponding to the $l$th pool. 
(Note that $\mathbf{z}_l$ is a vector of length $m_l$ and $\mathbf{x}_l$ is a vector of length $s$.)
We will estimate the support set $S_l$ of $\mathbf{x}_l$ as follows. 

\subsubsection{COMP Algorithm} First, we use the COMP algorithm to find a (superset) estimate $\widehat{S}_l$ of $S_l$. 
In particular, the COMP algorithm recovers $\widehat{S}_l$ from $\mathbf{z}_l$ given $\mathbf{A}_l$. 
Let $I_l = \{i\in [m_l]: (\mathbf{z}_l)_i = 0\}$, where $(\mathbf{z}_l)_i$ denotes the $i$th coordinate in $\mathbf{z}_l$. We denote by $\mathbf{x}^{*}_l$ the sub-vector of $\mathbf{x}_l$ restricted to the coordinates indexed by $\widehat{S}_l$; denote by $\mathbf{A}^{*}_l$ the sub-matrix of $\mathbf{A}_l$ restricted to the rows indexed by $[m_l]\setminus I_l$ and the columns indexed by $\widehat{S}_l$; and denote by $\mathbf{z}^{*}_l$ the sub-vector of $\mathbf{z}_l$ restricted to the coordinates indexed by $[m_l]\setminus I_l$. Let $m^{*}_l = |I_l|$ and $s^{*}_l = |\widehat{S}_l|$. In the next step, we will estimate the support set $S^{*}_l$ of $\mathbf{x}^{*}_l$ from $\mathbf{z}^{*}_l$, given $\mathbf{A}^{*}_l$. 

\subsubsection{MAP Decoding}
Given the estimate $\widehat{k}_l$ of $k_l$ (the number of nonzero coordinates in $\mathbf{x}^{*}_l$), let $k_{\min} = \max\{\widehat{k}_l-1,1\}$ and $k_{\max} = \min\{\widehat{k}_l+1,s^{*}_l\}$.  
For any $k_{\min}\leq k\leq k_{\max}$, and for any $k$-subset $T$ of $\widehat{S}_l$, we compute \[f(T)=\max_{\widehat{\mathbf{x}}^{*}_l:  \text{ support set of } \widehat{\mathbf{x}}^{*}_l \text{ is } T} p(\widehat{\mathbf{x}}^{*}_l|\mathbf{z}^{*}_l)\] by finding $\widehat{\mathbf{x}}^{*}_l$ with support set $T$ such that the conditional probability density of $\widehat{\mathbf{x}}^{*}_l$ given $\mathbf{z}^{*}_l$ is maximum. 
Maximizing $p(\widehat{\mathbf{x}}^{*}_l|\mathbf{z}^{*}_l)$ is equivalent to maximizing $p(\widehat{\mathbf{x}}^{*}_l) p(\mathbf{z}^{*}_l|\widehat{\mathbf{x}}^{*}_l) = \prod_{j\in [s^{*}_l]} p((\widehat{\mathbf{x}}^{*}_l)_j)\prod_{i\in [m^{*}_l]} p((\mathbf{z}^{*}_l)_i|\widehat{\mathbf{x}}^{*}_l)$, where $(\widehat{\mathbf{x}}^{*}_l)_j$ denotes the $j$th coordinate in $\widehat{\mathbf{x}}^{*}_l$. 
For any $\widehat{\mathbf{x}}^{*}_l$, $p((\widehat{\mathbf{x}}^{*}_l)_j) = (1-p)\delta((\widehat{\mathbf{x}}^{*}_l)_j) + p \times p_x((\widehat{\mathbf{x}}^{*}_l)_j)$, where $\delta(x)$ is the Dirac delta function, and $p((\mathbf{z}^{*}_l)_i|\widehat{\mathbf{x}}^{*}_l) = p_{\varepsilon}((\mathbf{z}^{*}_l)_i/((\mathbf{A}^{*}_l)_i \widehat{\mathbf{x}}^{*}_l))$, where $(\mathbf{A}^{*}_l)_i$ denotes the $i$th row of $\mathbf{A}^{*}_l$. 
Thus, if the sparsity parameter $p$ is known, $f(T)$ (for any $T$) can be approximated by solving a (potentially non-linear and/or non-convex) optimization problem (depending on the distributions $p_x$ and $p_{\varepsilon}$) numerically. 
(In our simulations, the ``fmincon'' function in MATLAB was used to compute an approximation of $f(T)$.) If $p$ is not known, $f(T)$ can be approximated similarly as above, except using the ML estimate $\widehat{p}$ everywhere, instead of $p$. 

\subsubsection{List Generation}\label{subsubsec:LG}
Let $f_{*} = \max_{T}f(T)$ where the maximization is over all $T$ defined as above. 
We find all $T$, say $T_1,T_2,\dots,T_{\ell}$, such that $f(T)\geq \alpha\cdot f_{*}$ for a given $0< \alpha\leq 1$, and 
use $T_1\cup T_2\cup \dots \cup T_{\ell}$ as the estimate of the support set $S^{*}_l$ of $\mathbf{x}^{*}_l$. 
Note that the larger is the threshold $\alpha$, the smaller will be the (conditional) average false negative rate and the larger will be the (conditional) average false positive rate.    

\section{2-STAMP: A Two-Stage Adaptive Mixed  Pooling}\label{subsec:TSSwithCP}
The proposed scheme in this section, which we refer to as the 2-STAMP scheme, is a generalization of the 2-STAP scheme. 
The first stage of the 2-STAMP scheme is the same as that in the 2-STAP scheme, but in the second stage we make measurements on mixtures of positive pools together, instead of making measurements on separate pools only. 
For the ease of exposition, we explain a special case of the 2-STAMP scheme when up to two pools can be mixed together. 
The same idea can be easily extended for mixing larger number of pools.

The main idea behind mixing pools in the second stage is as follows. Consider two positive pools that we expect to contain a relatively small number of nonzero coordinates. By mixing these pools together and sensing the mixed pool altogether (instead of sensing the pools individually), we can save a few measurements while maintaining the implementation/computational complexity of both the sensing and recovery algorithms affordable. 
However, the rest of the pools that are expected to contain a relatively large number of nonzero coordinates will be sensed individually, so as to avoid the sensing and/or recovery algorithms to become too complex implementation-wise or computationally. 

\subsection{Sensing Algorithm for Second Stage}
At the end of the first stage, suppose the noisy measurements $z^{(1)}_l = y^{(1)}_l\varepsilon^{(1)}_l$ for $l\in [q]$ are observed. 
Let $L = \{l\in [q]: z^{(1)}_l= 0\}$. 
Similarly as before, w.l.o.g., assume that $L = [q]\setminus [t]$ for some $0\leq t\leq q$. 
Let $\widehat{k}_1,\dots,\widehat{k}_t$ be our estimates of $k_1,\dots,k_t$, where $k_l$ is the number of nonzero coordinates in the $l$th pool. 
Assume, w.l.o.g., that $\widehat{k}_1\geq \widehat{k}_2\geq \dots \geq \widehat{k}_t$. 
For a given integer $\kappa\geq 1$, we partition the $t$ positive pools as follows: 
(i) find $\tau = \max\{l\in [t]: \widehat{k}_l>  \kappa\}$, and 
(ii) construct the partition $ \{1\},\dots,\{\tau\},\{\tau+1,\tau+2\},\{\tau+3,\tau+4\},\dots$. 
The last part is either $\{t-1,t\}$ or $\{t\}$ depending on whether $t-\tau$ is even or odd. 
Let $r = \lceil\frac{t+\tau}{2}\rceil$ be the number of parts in the partition. 
For the ease of exposition, we assume that the last part is $\{t-1,t\}$. 
We define $r$ mixed pools $\tilde{\mathbf{x}}_1,\dots, \tilde{\mathbf{x}}_{r}$, where $\tilde{\mathbf{x}}_l = \mathbf{x}_l$ for all $l\in [\tau]$ and $\tilde{\mathbf{x}}_l = [\mathbf{x}_{2l-\tau-1}^{\mathsf{T}},\mathbf{x}_{2l-\tau}^{\mathsf{T}}]^{\mathsf{T}}$ for all $l\in [r]\setminus [\tau]$. 
(Note that the $l$th pool has size $s$ or $2s$ for $l\in [\tau]$ or $l\in [r]\setminus [\tau]$, respectively.) 
Also, we define the $r$ noisy measurement vectors (corresponding to the first stage) for the mixed pools by $\tilde{\mathbf{z}}^{(1)}_1,\dots,\tilde{\mathbf{z}}^{(1)}_{r}$, where $\tilde{\mathbf{z}}^{(1)}_l = z^{(1)}_l$ for all $l\in [\tau]$ and $\tilde{\mathbf{z}}^{(1)}_l = [z^{(1)}_{2l-\tau-1},z^{(1)}_{2l-\tau}]^{\mathsf{T}}$ for all $l\in [r]\setminus [\tau]$.      

For each $l\in [\tau]$, we choose $m^{(2)}_l$ (the number of measurements on the $l$th mixed pool in the second stage) based on $\widehat{k}_{l}$ and we choose $\mathbf{A}^{(2)}_l$ (the sensing matrix for the $l$th mixed pool in the second stage) based on $m^{(2)}_l$, the same as that in the 2-STAP-II scheme. 
For each $l\in [r]\setminus [\tau]$, $m^{(2)}_l$ is chosen based on both $\widehat{k}_{2l-\tau-1}$ and $\widehat{k}_{2l-\tau}$, and $\mathbf{A}^{(2)}_l$ is chosen based on $m^{(2)}_l$, the same way as in the 2-STAP-II scheme, expect from an ensemble of binary matrices of size $m^{(2)}_l\times 2s$, instead of $m^{(2)}_l\times s$. 

\subsection{Recovery Algorithm for Second Stage}
The recovery algorithm in the second stage follows the same procedure as the one in the 2-STAP scheme, except that in this scheme we estimate the support set of each mixed pool (rather than estimating the support set of each pool individually). 
In particular, for each mixed pool $l\in [\tau]$ (which is essentially an individual pool), the recovery algorithm remains the same as that in the 2-STAP scheme. For each mixed pool $l\in [r]\setminus [\tau]$, we compute $f(T)$ for all subsets $T = T_1\cup T_2$ of $\widehat{S}_l$ (the support set estimated by the COMP algorithm for the $l$th mixed pool) such that $T_1$ is a subset of $\widehat{S}_l\cap \{1,\dots,s\}$ of size between $\max\{\widehat{k}_{2l-\tau-1}-1,1\}$ and $\min\{\widehat{k}_{2l-\tau-1}+1,|\widehat{S}_l\cap \{1,\dots,s\}|\}$, and $T_2$ is a subset of $\widehat{S}_l\cap \{s+1,\dots,2s\}$ of size between $\max\{\widehat{k}_{2l-\tau}-1,1\}$ and $\min\{\widehat{k}_{2l-\tau}+1,|\widehat{S}_l\cap \{s+1,\dots,2s\}|\}$. 

\section{Comparisons Between the Proposed Schemes and Tapestry}\label{sec:Comps}
The key differences between the proposed schemes and the Tapestry scheme are listed below: 
\begin{itemize}
    \item Tapestry is a single-stage scheme, and hence all measurements can be made in parallel. 
    The proposed schemes are, however, two-stage schemes; and notwithstanding that all measurements in each stage can be made in parallel, the measurements in the second stage can only be made after those in the first stage. 
    
    \item When compared to Tapestry, in the tested cases, the proposed schemes achieve a better tradeoff between the average number of measurements and the (conditional) average false negative/positive rate. 
    This comes from two facts: 
    (i) the sensing algorithm of the proposed schemes is more flexible than that of the Tapestry scheme. 
    This is because the total number of measurements in the latter can vary for different signal realizations, whereas the former is oblivious to different signal realizations and uses the same number of measurements always; and 
    (ii) the measurements in the proposed schemes are localized to small pools. This makes it possible to implement recovery algorithms that are carefully designed for the multiplicative noise model with reasonable computational complexity. In contrast, 
    the recovery algorithms discussed in the Tapestry scheme were borrowed from the CS literature where the noise model is assumed to be additive. 
    Unlike the recovery algorithms used for Tapestry, the recovery algorithm proposed in this work also takes into account the signal and noise distributions and the sparsity parameter or its estimate.  
    
    \item The sensing algorithm of the proposed schemes can potentially have a substantially lower computational and implementation complexity than that of the Tapestry scheme. 
    This follows from two facts: 
    (i) the total number of nonzero entries in the overall sensing matrix of the proposed schemes can be much smaller than that in the Tapestry scheme; and 
    (ii) the nonzero entries in each row of the sensing matrix in the Tapestry scheme are spread out everywhere, whereas the nonzero entries in the sensing matrix of the proposed schemes are localized in each row. 
    In particular, in the first stage each measurement is localized to a pool of consecutive coordinates, and the measurements in the second stage for each pool are over the coordinates in that pool only.
    
    
\end{itemize}

\section{Simulation Results}
In this section, we present our simulation results. 
As a case study, in these simulations, we have considered a population of $n=961$ people to be tested for COVID-19 and assumed that the prevalence is $p=0.01$. (In the simulations, we considered both cases where $p$ is known or unknown \emph{a priori}, and we did not observe any significant difference in the performance of either of the proposed schemes, 2-STAP and 2-STAMP.) 
For both the proposed schemes, we have considered pooling the population into $q=31$ pools, each of size $s=\frac{n}{q} = 31$, in the first stage. Three different values of number of infected people in the population ($k$), namely $k\in \{5,10,15\}$, have been considered. 
For every $k$, we performed $100$ Monte-Carlo simulations, where the statistical models used for viral load and measurement noise were obtained from~\cite{ghosh2020tapestry}. In particular, in each simulation trial, a signal of length $961$ was generated (independently from other trials) as follows: $k$ coordinates were randomly chosen to take a nonzero value from the interval $[1,1000]$ according to a (continuous) uniform distribution, and the remaining $n-k$ coordinates were all set to value zero. 
In the simulations, the measurement noise $\varepsilon$ was also assumed to have a log-normal distribution with parameters $\mu_{\varepsilon} = 0$ and $\sigma_{\varepsilon} = 0.1\ln(1.95)$.


Table~\ref{tab:All} summarizes our results for the proposed 2-STAP (both variants) and 2-STAMP schemes and the results for the Tapestry scheme for the same problem model (i.e., the same population size and the same viral load and noise distributions) where each measurement is made on a pool of $31$ people (see~\cite{ghosh2020tapestry}). 
In this table, $m_{\min}$, $m_{\max}$, $m_{\mathrm{std}}$, and $m_{\mathrm{ave}}$ represent the minimum, maximum, standard deviation, and the average of the number of measurements used in $100$ simulations, respectively. The sensitivity and specificity results are rounded to two decimal places, for fair comparison with the results reported in~\cite{ghosh2020tapestry}. More detailed results for the 2-STAP-I, 2-STAP-II, and 2-STAMP schemes are presented in Tables~\ref{tab:I},~\ref{tab:II}, and~\ref{tab:III}, respectively. In these tables, (i) $\alpha$ is the threshold used in the list generation step of the recovery algorithm (see Section~\ref{subsubsec:LG}), which controls the trade-off between sensitivity and specificity; and (ii) the \emph{conditional negative (or positive) predictive value} is the average of the ratio of the number of true negatives (or positives) to the total number of true and false negatives (or positives). Note that the negative predictive value represents the probability that a person is truly non-infected given that the recovery algorithm identifies them as non-infected, and the positive predictive value represents the probability that a person is truly infected given that the recovery algorithm identifies them as infected.     


\begin{table}[]
\begin{center}
\caption{Performance results for the Tapestry scheme~\cite{ghosh2020tapestry} and the proposed 2-STAP and 2-STAMP schemes (sensitivity and specificity results are rounded to two decimal places)}\label{tab:All}
\begin{tabular}{|c|c|c|c|c|c|c|c|}
\hline
$k$ & $m_{\min}$ & $m_{\max}$ & $m_{\mathrm{std}}$ & $m_{\mathrm{ave}}$ & Pooling Scheme & \begin{tabular}[c]{@{}c@{}}Conditional \\ Sensitivity\end{tabular} & \begin{tabular}[c]{@{}c@{}}Conditional\\ Specificity\end{tabular} \\ \hline
\multirow{3}{*}{5} & 93 & 93 & 0 & 93 & \begin{tabular}[c]{@{}c@{}}Tapestry + COMP + NN-LASSO \cite{ghosh2020tapestry}\\ Tapestry + COMP + NN-OMP \cite{ghosh2020tapestry}\\ Tapestry + COMP + SBL \cite{ghosh2020tapestry}\end{tabular} & \begin{tabular}[c]{@{}c@{}}1.00\\ 1.00\\ 1.00\end{tabular} & \begin{tabular}[c]{@{}c@{}}1.00\\ 1.00\\ 1.00\end{tabular} \\ \cline{2-8} 
 & 49 & 61 & 2.94 & 59.08 & 2-STAP-I  & 1.00 & 1.00 \\ \cline{2-8}
 & 47 & 57 & 2.26 & 54.55 & 2-STAP-II  & 1.00 & 1.00 \\ \cline{2-8} 
 & 46 & 55 & 2.25 & 52.56 & 2-STAMP & 1.00 & 1.00 \\ \hline
\multirow{3}{*}{10} & 93 & 93 & 0 & 93 & \begin{tabular}[c]{@{}c@{}}Tapestry + COMP + NN-LASSO \cite{ghosh2020tapestry}\\ Tapestry + COMP + NN-OMP \cite{ghosh2020tapestry}\\ Tapestry + COMP + SBL \cite{ghosh2020tapestry} \end{tabular} & \begin{tabular}[c]{@{}c@{}}0.98\\ 0.96\\ 0.99\end{tabular} & \begin{tabular}[c]{@{}c@{}}0.99\\ 1.00\\ 0.99\end{tabular} \\ \cline{2-8} 
 & 73 & 91 & 5.54 & 82.96 & 2-STAP-I  & 1.00 & 1.00 \\ \cline{2-8}
 & 66 & 81 & 4.18 & 74.98 & 2-STAP-II  & 0.99 & 1.00 \\ \cline{2-8} 
 & 63 & 76 & 3.74 & 70.86 & 2-STAMP  & 1.00 & 1.00 \\ \hline
\multirow{3}{*}{15} & 93 & 93 & 0 & 93 & \begin{tabular}[c]{@{}c@{}}Tapestry + COMP + NN-LASSO \cite{ghosh2020tapestry}\\ Tapestry + COMP + NN-OMP \cite{ghosh2020tapestry}\\ Tapestry + COMP + SBL \cite{ghosh2020tapestry}\end{tabular} & \begin{tabular}[c]{@{}c@{}}0.94\\ 0.86\\ 0.98\end{tabular} & \begin{tabular}[c]{@{}c@{}}0.97\\ 0.99\\ 0.97\end{tabular} \\ \cline{2-8} 
 & 85 & 121 & 7.31 & 103.30 & 2-STAP-I  & 0.98 & 0.99 \\ \cline{2-8}
 & 78 & 106 & 5.65 & 92.66 & 2-STAP-II  & 0.98 & 0.99 \\ \cline{2-8} 
 & 74 & 99 & 5.02 & 86.85 & 2-STAMP & 0.99 & 0.99 \\ \hline
\end{tabular}
\end{center}
\end{table}

\begin{table}[t]
\begin{center}
\caption{Detailed performance results for the 2-STAP-I scheme}\label{tab:I}
\begin{tabular}{|c|c|c|c|c|c|c|c|c|c|}
\hline
$k$ & $m_{\min}$ & $m_{\max}$ & $m_{\mathrm{std}}$ & $m_{\mathrm{ave}}$ & $\alpha$ & \begin{tabular}[c]{@{}c@{}} Conditional\\[-0.125cm] Sensitivity\end{tabular} & \begin{tabular}[c]{@{}c@{}} Conditional\\[-0.125cm] Specificity\end{tabular} & \begin{tabular}[c]{@{}c@{}} Conditional Negative \\[-0.125cm] Predictive Value \end{tabular} & \begin{tabular}[c]{@{}c@{}} Conditional Positive\\[-0.125cm] Predictive Value \end{tabular} \\ \hline
5 & 49 & 61 & 2.94 & 59.08 & \begin{tabular}[c]{@{}c@{}} 0.943 \\ 0.926 \\ 0.893 \end{tabular} & \begin{tabular}[c]{@{}c@{}} 0.9920 \\ 0.9960 \\ 0.9980 \end{tabular} & \begin{tabular}[c]{@{}c@{}} 0.9995 \\ 0.9994 \\ 0.9993 \end{tabular} & \begin{tabular}[c]{@{}c@{}} 1.0000 \\ 1.0000 \\ 1.0000 \end{tabular} & \begin{tabular}[c]{@{}c@{}} 0.9495 \\ 0.9492 \\ 0.9374 \end{tabular} \\ \hline
{ 10} & 73 & 91 & 5.54 & 82.96 & \begin{tabular}[c]{@{}c@{}} 0.943 \\ 0.909 \\ 0.885 \end{tabular} & \begin{tabular}[c]{@{}c@{}}  0.9810 \\ 0.9930 \\ 0.9950 \end{tabular} & \begin{tabular}[c]{@{}c@{}} 0.9988 \\ 0.9982 \\ 0.9978 \end{tabular} & \begin{tabular}[c]{@{}c@{}} 0.9998 \\ 0.9999 \\ 0.9999 \end{tabular} & \begin{tabular}[c]{@{}c@{}} 0.9225 \\ 0.8980 \\ 0.8853 \end{tabular} \\ \hline
15 & 85 & 121 & 7.30 & 103.30 & \begin{tabular}[c]{@{}c@{}} 0.847 \\ 0.758 \\ 0.649 \end{tabular} & \begin{tabular}[c]{@{}c@{}} 0.9820 \\ 0.9893 \\ 0.9940 \end{tabular} & \begin{tabular}[c]{@{}c@{}} 0.9939 \\ 0.9885 \\ 0.9817 \end{tabular} & \begin{tabular}[c]{@{}c@{}} 0.9997 \\ 0.9998 \\ 0.9999 \end{tabular} & \begin{tabular}[c]{@{}c@{}} 0.7709 \\ 0.6475 \\ 0.5290 \end{tabular} \\ \hline
\end{tabular}
\end{center}
\end{table}

\begin{table}[t]
\begin{center}
\caption{Detailed performance results for the 2-STAP-II scheme}\label{tab:II}
\begin{tabular}{|c|c|c|c|c|c|c|c|c|c|}
\hline
$k$ & $m_{\min}$ & $m_{\max}$ & $m_{\mathrm{std}}$ & $m_{\mathrm{ave}}$ & $\alpha$ & \begin{tabular}[c]{@{}c@{}} Conditional\\[-0.125cm] Sensitivity\end{tabular} & \begin{tabular}[c]{@{}c@{}} Conditional\\[-0.125cm] Specificity\end{tabular} & \begin{tabular}[c]{@{}c@{}} Conditional Negative \\[-0.125cm] Predictive Value \end{tabular} & \begin{tabular}[c]{@{}c@{}} Conditional Positive\\[-0.125cm] Predictive Value \end{tabular} \\ \hline
5 & 47 & 57 & 2.26 & 54.55 & \begin{tabular}[c]{@{}c@{}} 0.813\\ 0.794\\ 0.787\end{tabular} & \begin{tabular}[c]{@{}c@{}}0.9920\\ 0.9940\\ 0.9980\end{tabular} & \begin{tabular}[c]{@{}c@{}}0.9986\\ 0.9983\\ 0.9982\end{tabular} & \begin{tabular}[c]{@{}c@{}} 0.9999 \\ 1.0000 \\ 1.0000 \end{tabular} & \begin{tabular}[c]{@{}c@{}} 0.8259 \\ 0.8026 \\ 0.7953 \end{tabular} \\ \hline
10 & 66 & 81 & 4.18 & 74.98 & \begin{tabular}[c]{@{}c@{}} 0.943\\ 0.794\\ 0.730 \end{tabular} & \begin{tabular}[c]{@{}c@{}} 0.9810 \\ 0.9930\\ 0.9950 \end{tabular} & \begin{tabular}[c]{@{}c@{}}  0.9987\\ 0.9954\\ 0.9927\end{tabular} & \begin{tabular}[c]{@{}c@{}} 0.9998 \\ 0.9999 \\ 0.9999 \end{tabular} & \begin{tabular}[c]{@{}c@{}} 0.9074 \\ 0.7471 \\ 0.6559 \end{tabular}\\ \hline
15 & 78 & 106 & 5.65 & 92.66 & \begin{tabular}[c]{@{}c@{}} 0.787\\ 0.671\\ 0.515\end{tabular} & \begin{tabular}[c]{@{}c@{}}0.9820\\ 0.9893\\ 0.9940\end{tabular} & \begin{tabular}[c]{@{}c@{}}0.9903\\ 0.9812\\ 0.9724\end{tabular} & \begin{tabular}[c]{@{}c@{}} 0.9997 \\ 0.9998 \\ 0.9999 \end{tabular} & \begin{tabular}[c]{@{}c@{}} 0.6699 \\ 0.5052 \\ 0.4109 \end{tabular} \\ \hline
\end{tabular}
\end{center}
\end{table}

\begin{table}[t]
\begin{center}
\caption{Detailed performance results for the 2-STAMP scheme}\label{tab:III}
\begin{tabular}{|c|c|c|c|c|c|c|c|c|c|}
\hline
$k$ & $m_{\min}$ & $m_{\max}$ & $m_{\mathrm{std}}$ & $m_{\mathrm{ave}}$ & $\alpha$ & \begin{tabular}[c]{@{}c@{}} Conditional\\[-0.125cm] Sensitivity\end{tabular} & \begin{tabular}[c]{@{}c@{}} Conditional\\[-0.125cm] Specificity\end{tabular} & \begin{tabular}[c]{@{}c@{}} Conditional Negative \\[-0.125cm] Predictive Value \end{tabular} & \begin{tabular}[c]{@{}c@{}} Conditional Positive\\[-0.125cm] Predictive Value \end{tabular} \\ \hline
5 & 46 & 55 & 2.25 & 52.56 & \begin{tabular}[c]{@{}c@{}} 0.893 \\ 0.877 \\ 0.847 \end{tabular} & \begin{tabular}[c]{@{}c@{}} 0.9920 \\ 0.9940 \\ 0.9980 \end{tabular} & \begin{tabular}[c]{@{}c@{}} 0.9989 \\ 0.9986 \\ 0.9978 \end{tabular} & \begin{tabular}[c]{@{}c@{}} 1.0000 \\ 1.0000 \\ 1.0000 \end{tabular} & \begin{tabular}[c]{@{}c@{}} 0.8698 \\ 0.8473 \\ 0.7832 \end{tabular}\\ \hline
{10} & 63 & 76 & 3.74 & {70.86} & \begin{tabular}[c]{@{}c@{}} 0.980 \\ 0.935 \\ 0.885 \end{tabular} & \begin{tabular}[c]{@{}c@{}} {0.9800}\\ {0.9930}\\ {0.9950}\end{tabular} & \begin{tabular}[c]{@{}c@{}}{0.9990}\\ {0.9981}\\ {0.9962}\end{tabular} & \begin{tabular}[c]{@{}c@{}} 0.9998 \\ 0.9999 \\ 0.9999 \end{tabular} & \begin{tabular}[c]{@{}c@{}} 0.9324 \\ 0.8830 \\ 0.7955 \end{tabular} \\ \hline
15 & 74 & 99 & 5.02 & 86.85 & \begin{tabular}[c]{@{}c@{}} 0.909 \\ 0.877 \\ 0.840 \end{tabular} & \begin{tabular}[c]{@{}c@{}} 0.9813 \\ 0.9900 \\ 0.9947 \end{tabular} & \begin{tabular}[c]{@{}c@{}} 0.9945 \\ 0.9916 \\ 0.9847 \end{tabular} & \begin{tabular}[c]{@{}c@{}} 1.0000 \\ 1.0000 \\ 1.0000 \end{tabular} & \begin{tabular}[c]{@{}c@{}} 0.9010 \\ 0.8474 \\ 0.7736 \end{tabular} \\ \hline
\end{tabular}
\end{center}
\end{table}

Comparing the results of Tapestry and the two variants of 2-STAP in Table~\ref{tab:All}, it can be seen that for $k\in \{5,10\}$, 2-STAP-I requires smaller number of measurements on the average for the same (or even higher) sensitivity and specificity (see also Table~\ref{tab:I}). For $k=15$, 2-STAP-I uses about $10$ more measurements than Tapestry on the average, but it achieves a substantially higher specificity by about $2\%$ for almost the same sensitivity (see also Table~\ref{tab:I}). It can also be seen that for all $k\in \{5,10,15\}$, 2-STAP-II can provide higher sensitivity and higher specificity than Tapestry with even smaller (average) number of measurements. 
For instance, for the case of $k=10$ infected people, with an average number of measurements about $75$, 2-STAP-II can achieve a sensitivity of $99.30\%$ and a specificity of $99.54\%$, see Table~\ref{tab:II}. 
However, using Tapestry, for $k=10$, one can achieve a sensitivity and a specificity between $98.50\%$ and $99.49\%$ with $93$ measurements (about $20\%$ more measurements than that in 2-STAP-II). 
These improvements in the performance are mainly due to the fact that 2-STAP is an adaptive scheme (although with a very small degree of adaptivity, i.e., using only one round of feedback), whereas Tapestry is a non-adaptive scheme. 
In particular, identifying all negative pools (which contain a relatively large fraction of population for sufficiently low prevalence) at the end of the first stage and using a relatively small number of additional measurements only for each positive pool in the second stage enable us to achieve a better trade-off between average number of measurements, sensitivity, and specificity.  

As can be seen in Table~\ref{tab:All}, 2-STAMP can achieve a sensitivity and a specificity higher than those attainable with 2-STAP-I and 2-STAP-II, with even smaller average number of measurements. 
For instance, for $k=10$, with only about $71$ ($<75$ in 2-STAP-II) measurements on average, 2-STAMP can achieve a sensitivity of $99.30\%$ (the same as that in 2-STAP-II) and a specificity of $99.81\%$ ($>99.54\%$ in 2-STAP-II), see Table~\ref{tab:II}.  
The advantage of 2-STAMP over both variants of 2-STAP comes from the saving in the number of measurements in the second stage. 
In particular, in 2-STAMP, mixing small groups (namely, groups of size two) of pools with small number of infected people gives rise to an opportunity for making a smaller number of measurements on the mixed super-pool (as compared to the total number of measurements used in 2-STAP for all pools in the mix) without compensating the overall accuracy.

\bibliographystyle{IEEEtran}
\bibliography{grouptesting,0-Bibliography,COVID-19}

\appendices

\section{Detailed Explanation about Multiplicative Noise Model}\label{app:0}
The multiplicative noise model is inspired by the current RT-qPCR technology for COVID-19 testing. 
To keep the discussion simple, suppose that an individual person is to be tested for COVID-19 by using this technology. 
The sample collected from the person to be tested is dispersed into a liquid medium, and the reverse transcription (RT) process is applied to convert the RNA molecules of the SARS-CoV-2 virus (the coronavirus that causes COVID-19) in the liquid (if the person is infected) into cDNA. 
Followed by adding primers that are complementary to the cDNA of the viral genome, these primers attach themselves to the cDNA of the viral genome, and together they undergo an exponential amplification process by the RT-qPCR machine \cite{ghosh2020tapestry}. 
This process consists of a maximum of $C_{\max}$ cycles. 
The output of the RT-qPCR process is the cycle count $C$ after which the concentration of DNA 
exceeds a pre-specified threshold $D_{\min}$ or $C_{\max}$.
The thresholds $D_{\min}$ and $C_{\max}$ are often chosen so that: 
(i) if a person is not infected, the DNA concentration does not exceed $D_{\min}$ over the course of $C_{\max}$ cycles, and 
(ii) if a person is infected, the DNA concentration exceeds $D_{\min}$ at some point, say cycle $C$, over the course of $C_{\max}$ cycles. 
Note that for a fixed $D_{\min}$, the larger is the viral load of an infected person, the smaller $C$ will be. 
Ideally, the concentration of DNA molecules is doubled in every cycle, i.e., after $C$ cycles the concentration of DNA molecules is $x 2^C$, where $x$ is the viral load of the person to be tested. 
In reality, however, the amplification process may not be ideal. 
To reflect the randomness in the process, we use the same model as the one suggested in~\cite{ghosh2020tapestry} and assume that the concentration of DNA molecules after $C$ cycles is given by $x b^{C+\Delta}$ for some positive constant $b$ (close to $2$), 
where $x$ is the viral load of the person to be tested, and $\Delta$ is a Gaussian random variable with mean zero and variance $\sigma^2_{\Delta}$.
The multiplicative term $\varepsilon = b^{\Delta}$ can be viewed as the noise in the amplification process. 
The random variable $\varepsilon$ has a log-normal distribution with parameters $\mu_{\varepsilon}=0$ and $\sigma_{\varepsilon} = \sigma_{\Delta}\ln b$. 
Note that the closer are $b$ to $2$ and $\sigma_{\Delta}$ to $0$, the weaker will be the noise and the closer will be the process to ideal. 
For any $x>0$, let $C_x$ be the number of cycles it takes for the concentration of DNA molecules to be approximately equal to $D_{\min}$. 
That is, $x b^{C_x+\Delta}\approx D_{\min}$. 
The cycle count measurement $C_x$ can be converted to an equivalent measurement $z$ given by $z = D_{\min} b^{-C_x}$.
It can then be seen that $z = x\varepsilon$ gives us the measurement model with multiplicative noise in \eqref{eqn:multiplicativenoise}.

\section{Other Recovery Algorithms in~\cite{ghosh2020tapestry}}\label{app:I}

Recall the index set $I = \{i\in [m]: z_i=0\}$ of zero measurements and the superset estimate $\widehat{S}(\mathbf{x})$ of $S(\mathbf{x})$ from the COMP algorithm. For simplifying the notation, denote $S(\mathbf{x})$ and $\widehat{S}(\mathbf{x})$ by $S$ and $\widehat{S}$, respectively. 
Also, denote by $\mathbf{x}^{*}$ the sub-vector of $\mathbf{x}$ restricted to the coordinates indexed by $\widehat{S}$; denote by $\mathbf{A}^{*}$ the sub-matrix of $\mathbf{A}$ restricted to the rows indexed by $[m]\setminus I$ and the columns indexed by $\widehat{S}$; and denote by $\mathbf{z}^{*}$ the sub-vector of $\mathbf{z}$ restricted to the coordinates indexed by $[m]\setminus I$. Let $m^{*} = |I|$ and $n^{*} = |\widehat{S}|$. Note that $\mathbf{x}^{*}$ is a vector of length $n^{*}$, $\mathbf{A}^{*}$ is an $m^{*}\times n^{*}$ matrix, and $\mathbf{z}^{*}$ is a vector of length $m^{*}$. 

The rest of the recovery algorithms discussed in~\cite{ghosh2020tapestry} consist of two phases. In the first phase, all of these algorithms use COMP for an initial signal-support recovery and reduce the instance $(\mathbf{x},\mathbf{A},\mathbf{z})$ to the instance $(\mathbf{x}^{*},\mathbf{A}^{*},\mathbf{z}^{*})$, and in the second phase, each of these algorithms employs a different technique for signal recovery. 
The second phase of these algorithms are briefly described as follows. 

For simplifying the notation, in the following we omit the superscript ``$*$'', and denote $\mathbf{x}^{*},\mathbf{A}^{*},\mathbf{z}^{*},m^{*},n^{*}$ by $\mathbf{x},\mathbf{A},\mathbf{z},m,n$, respectively.    

\subsubsection*{Non-negative LASSO (NN-LASSO)}
NN-LASSO is a sparse signal recovery technique to compute an estimate $\widehat{\mathbf{x}}$ of $\mathbf{x}$ by trying to minimize the $L_2$-norm of the residual (additive) error vector $\mathbf{z}-\mathbf{A}\widehat{\mathbf{x}}$, 
i.e., $\|\mathbf{z}-\mathbf{A}\widehat{\mathbf{x}}\|_2$, subject to 
(i) a sparsity constraint on $\widehat{\mathbf{x}}$ -- imposed by an upper bound on the $L_1$-norm of $\widehat{\mathbf{x}}$, 
i.e., $\|\widehat{\mathbf{x}}\|_1\leq \lambda$, and 
(ii) a non-negativity constraint on all coordinates in $\widehat{\mathbf{x}}$. 
The choice of the threshold $\lambda$ depends on the sensing matrix $\mathbf{A}$, the noisy measurement vector $\mathbf{z}$, and the distribution of the noise $\varepsilon$ (for details, see~\cite{ghosh2020tapestry}).  

\subsubsection*{Non-negative Orthogonal Matching Pursuit (NN-OMP)}
NN-OMP is an iterative greedy technique for non-negative sparse signal recovery that finds an estimate $\widehat{\mathbf{x}}$ of $\mathbf{x}$ by seeking to minimize the $L_0$-norm of $\widehat{\mathbf{x}}$, 
i.e., $\|\widehat{\mathbf{x}}\|_0$, 
subject to (i) a constraint on the $L_2$-norm of the residual (additive) error vector $\mathbf{z}-\mathbf{A}\widehat{\mathbf{x}}$, i.e., $\|\mathbf{z}-\mathbf{A}\widehat{\mathbf{x}}\|_2\leq \epsilon$, and 
(ii) a non-negativity constraint on all coordinates in $\widehat{\mathbf{x}}$. 
Followed by initializing $\mathbf{r}^{(0)}$ by $\mathbf{z}$ and $\widehat{S}^{(0)}$ by the empty set, 
in each iteration $l$, 
the algorithm first updates the support set $\widehat{S}^{(l)} = \widehat{S}^{(l-1)}\cup \{j^{(l)}\}$, 
where the $j^{(l)}$th column of $\mathbf{A}$ has the maximum correlation with the residual error vector $\mathbf{r}^{(l-1)}$. 
That is, $j^{(l)} = \argmax_{j\in [n]} \mathbf{A}_{j}^{\mathsf{T}}\mathbf{r}^{(l-1)}$, where $\mathbf{A}_j$ denotes the $j$th column of $\mathbf{A}$. 
Next, the algorithm updates the residual error vector $\mathbf{r}^{(l)} = \mathbf{z} - \mathbf{A}\widehat{\mathbf{x}}$, 
where $\widehat{\mathbf{x}}$ minimizes $\|\mathbf{z}-\mathbf{A}\widehat{\mathbf{x}}\|_2$ subject to $\widehat{x}_j\geq 0$ for all $j\in \widehat{S}^{(l)}$, and $\widehat{x}_j=0$ for all $j\in [n]\setminus\widehat{S}^{(l)}$. 
Once the stopping criterion $\|\mathbf{z}-\mathbf{A}\widehat{\mathbf{x}}\|_2\leq \epsilon$ is satisfied, 
the algorithm terminates and 
returns $\widehat{\mathbf{x}}$ as an estimate of $\mathbf{x}$. 
The choice of the threshold $\epsilon$ depends on $\mathbf{A}$, $\mathbf{z}$, and the distribution of $\varepsilon$ (see~\cite{ghosh2020tapestry}). 

\subsubsection*{Sparse Bayesian Learning (SBL)}
In SBL, for an estimate $\widehat{\mathbf{x}}$ of $\mathbf{x}$, 
the $j$th coordinate in $\widehat{\mathbf{x}}$ is assumed to be an independent Gaussian random variable with mean zero and variance $\sigma^2_j$, and
the coordinates of the residual (additive) error vector $\mathbf{z}-\mathbf{A}\widehat{\mathbf{x}}$ are assumed to be independent Gaussian random variables with mean zero and variance $\sigma^2$.
Under these assumptions, 
the likelihood of $\widehat{\mathbf{x}}$ is given by 
$p(\widehat{\mathbf{x}};\{\sigma_j\}) = \prod_{j\in [n]} (2\pi\sigma^2_j)^{-\frac{1}{2}}\exp(-\frac{1}{2}\sigma_j^{-2}\widehat{x}_j^2)$, and 
the conditional likelihood of $\mathbf{z}$ given $\widehat{\mathbf{x}}$ is given by 
$p(\mathbf{z}|\widehat{\mathbf{x}};\sigma) = (2\pi\sigma^2)^{-\frac{m}{2}}\exp(-\frac{1}{2}\sigma^{-2}\|\mathbf{z}-\mathbf{A}\widehat{\mathbf{x}}\|^2_2)$. 
Let $\mathbf{I}$ be an $m\times m$ identity matrix, and let 
$\mathbf{\Sigma}_z = \sigma^2 \mathbf{I} + \mathbf{A}\hspace{1pt} \mathrm{diag}(\{\sigma^2_j\}) \mathbf{A}^{\mathsf{T}}$, where $\mathrm{diag}(\{v_j\})$ is a square diagonal matrix such that the $(j,j)$th entry is $v_j$.  
Marginalizing the joint distribution $p(\mathbf{z},\widehat{\mathbf{x}};\{\sigma_j\},\sigma) = p(\widehat{\mathbf{x}};\{\sigma_j\}) p(\mathbf{z}|\widehat{\mathbf{x}};\sigma)$, 
the likelihood of $\mathbf{z}$ is given by
$p(\mathbf{z};\{\sigma_j\},\sigma) = (2\pi)^{-\frac{m}{2}} |\mathbf{\Sigma}_z|^{-\frac{1}{2}}\exp(-\frac{1}{2}{\mathbf{z}}^{\mathsf{T}} \mathbf{\Sigma}_z^{-1}\mathbf{z})$.
SBL uses the Expectation-Maximization (EM) algorithm, which is an iterative technique, to find $\{\sigma_j\}$ and $\sigma$ that maximize $p(\mathbf{z};\{\sigma_j\},\sigma)$. 
Followed by initializing $\{\sigma_j\}$ by $\{\sigma_j^{(0)}\}$ and $\sigma$ by $\sigma^{(0)}$, 
in each iteration $l$, 
the EM algorithm updates $\sigma_j^{(l)}$ and $\sigma^{(l)}$ as follows:
\[\sigma_j^{(l)} = \left((\mathbf{\Sigma}_x)_{j,j} + \widehat{x}^2_j\right)^{\frac{1}{2}}\] 
\begin{dmath*}\sigma^{(l)} = m^{-\frac{1}{2}}\left( {\|\mathbf{z}-\mathbf{A}\widehat{\mathbf{x}}\|_2^2} + (\sigma^{(l-1)})^{2} \sum_{j\in [n]} \left(1- (\sigma^{(l-1)}_j)^{-2} (\mathbf{\Sigma}_x)_{j,j} \right)  \right)^{\frac{1}{2}}\end{dmath*} 
where $(\mathbf{\Sigma}_x)_{j,j}$ is the $(j,j)$th entry in the matrix $\mathbf{\Sigma}_x = ((\sigma^{(l-1)})^{-2}\mathbf{A}^{\mathsf{T}}\mathbf{A}+\mathrm{diag}(\{(\sigma^{(l-1)}_j)^{-2}\}))^{-1}$ and $\widehat{\mathbf{x}} = (\sigma^{(l-1)})^{-2}\mathbf{\Sigma}_x\mathbf{A}^{\mathsf{T}}\mathbf{z}$. 
The iterations continue until convergence of $\widehat{\mathbf{x}}$. 
(The EM algorithm is guaranteed to converge to a fixed point, which may or may not be a local optimum.) 
Upon convergence, the algorithm terminates and returns $\widehat{\mathbf{x}}$. Since $\widehat{\mathbf{x}}$ may have some negative coordinates, as suggested in~\cite{ghosh2020tapestry} one can use 
$[\max\{\widehat{x}_1,0\},\dots,\max\{\widehat{x}_n,0\}]^{\mathsf{T}}$ as an estimate of $\mathbf{x} = [x_1,\dots,x_n]^{\mathsf{T}}$. 
A smarter yet more elaborate approach to impose the non-negativity constraint as part of the optimization problem is to use algorithms such as Rectified SBL~\cite{NFASLR2018} that assume the coordinates in $\widehat{\mathbf{x}}$ follow a different distribution than the Gaussian distribution used in SBL.    


\section{Number of Measurements and Sensing Matrices Used in Simulations}\label{app:II}

\begin{figure}[h]
	\centering
	\includegraphics[width=6.5in]{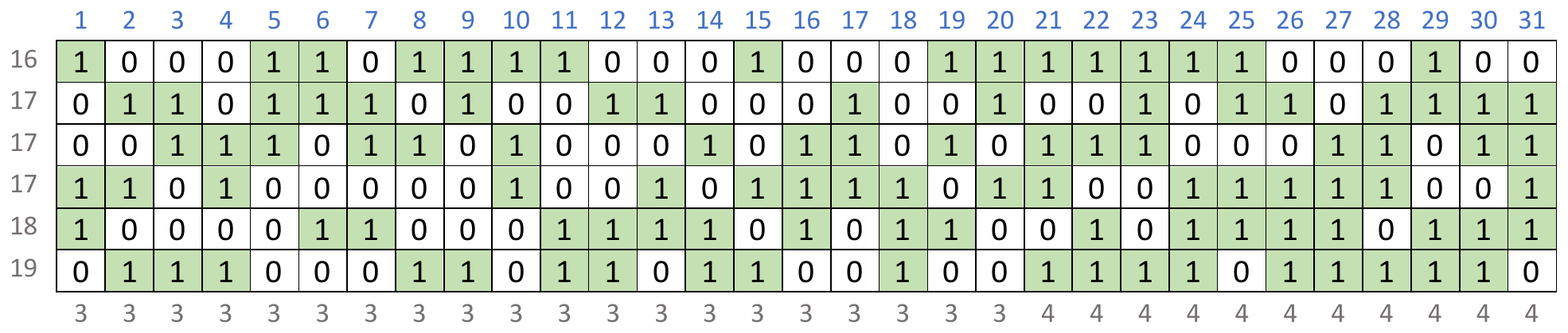}\vspace{-0.45cm}
	\caption{The sensing matrix with $6$ rows and $31$ columns used in the simulations for the 2-STAP-I scheme. (The number of 1's in each row is shown in front of the row, and the number of 1's in each column is shown below the column.)}\label{fig:MatricesI}
\end{figure}

For the 2-STAP-I scheme, for each positive pool $l$, $m^{(2)}_l=6$ measurements were used in the second stage, and the pooling matrix used in the simulations can be found in Fig.~\ref{fig:MatricesI}.

\begin{figure}[h]
	\centering
	\includegraphics[width=6.5in]{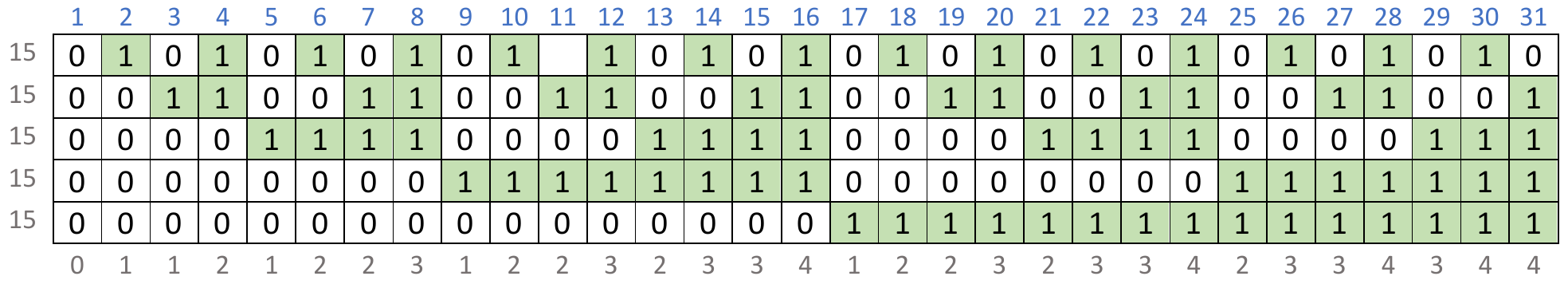}\vspace{-0.45cm}
	\begin{center}
	    {\footnotesize (a) $5\times 31$ sensing matrix}\vspace{0.45cm}
	\end{center}
	\includegraphics[width=6.5in]{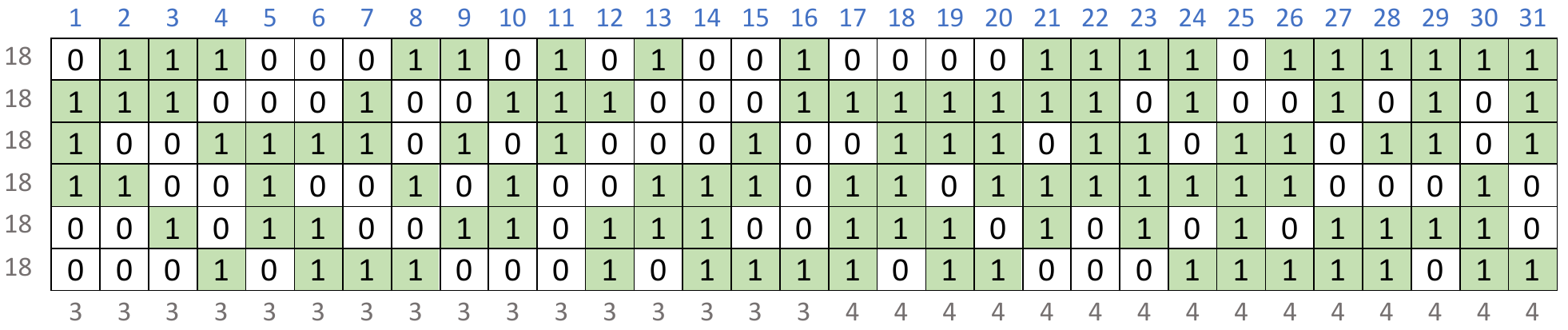}\vspace{-0.45cm}
	\begin{center}
	    {\footnotesize (b) $6\times 31$ sensing matrix}\vspace{0.45cm}
	\end{center}
	\includegraphics[width=6.5in]{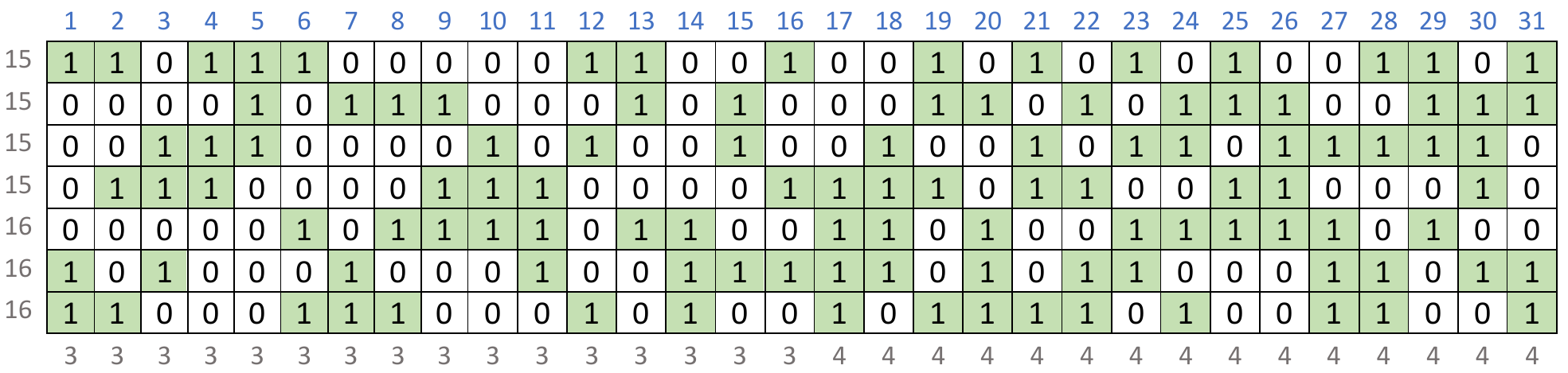}\vspace{-0.45cm}
	\begin{center}
	    {\footnotesize (c) $7\times 31$ sensing matrix}\vspace{0.45cm}
	\end{center}
	\includegraphics[width=6.5in]{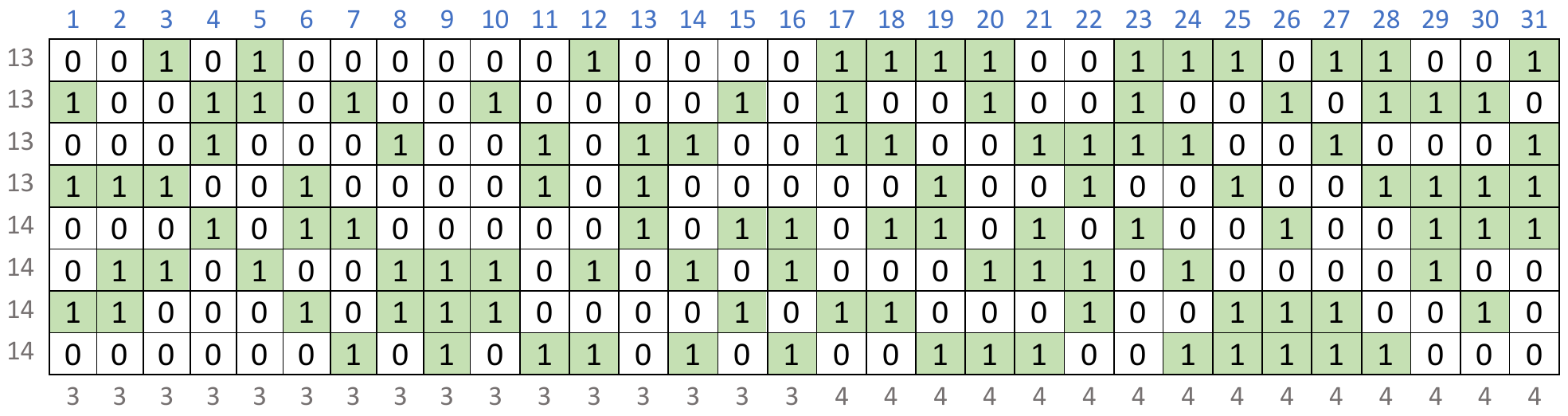}\vspace{-0.45cm}
	\begin{center}
	    {\footnotesize (d) $8\times 31$ sensing matrix}
	\end{center}
	\caption{The sensing matrices with $5,6,7,8$ rows and $31$ columns used in the simulations for the 2-STAP-II scheme. (The weight of each row is shown in front of the row, and the weight of each column is shown below the column.)}\label{fig:MatricesII}
\end{figure}

For the 2-STAP-II scheme, the number of measurements $m^{(2)}_l$ used in the second stage for each positive pool $l$ was chosen as follows (depending on the estimate $\widehat{k}_l$ of the number of infected people in pool $l$): 
\[m^{(2)}_l = 
\begin{cases}
5, & \widehat{k}_l = 1,\\
6, & \widehat{k}_l = 2,\\
7, & \widehat{k}_l = 3,\\
8, & \widehat{k}_l \geq 4.
\end{cases}
\] For each $m^{(2)}_l$, the column and row weight distributions of the $m^{(2)}_l\times 31$ sensing matrix $\mathbf{A}^{(2)}_l$ used in the simulations are described below (see Fig.~\ref{fig:MatricesII}): 
\begin{itemize}
    \item $m^{(2)}_l=5$: 
    \begin{itemize}
        \item $1,5,10,10,5$ columns of weights $0,1,2,3,4$, respectively;
        \item $5$ rows of weight $15$.
    \end{itemize}
    \item $m^{(2)}_l=6$: 
    \begin{itemize}
        \item $16,15$ columns of weights $3,4$, respectively;
        \item $6$ rows of weight $18$.
    \end{itemize}
    \item $m^{(2)}_l=7$: 
    \begin{itemize}
        \item $16,15$ columns of weights $3,4$, respectively;
        \item $4,3$ rows of weights $15,16$, respectively.
    \end{itemize}
    \item $m^{(2)}_l=8$: 
    \begin{itemize}
        \item $16,15$ columns of weights $3,4$, respectively;
        \item $4,4$ rows of weights $13,14$, respectively.
    \end{itemize}
\end{itemize}

\begin{figure}[h]
	\centering
	\includegraphics[width=6.5in]{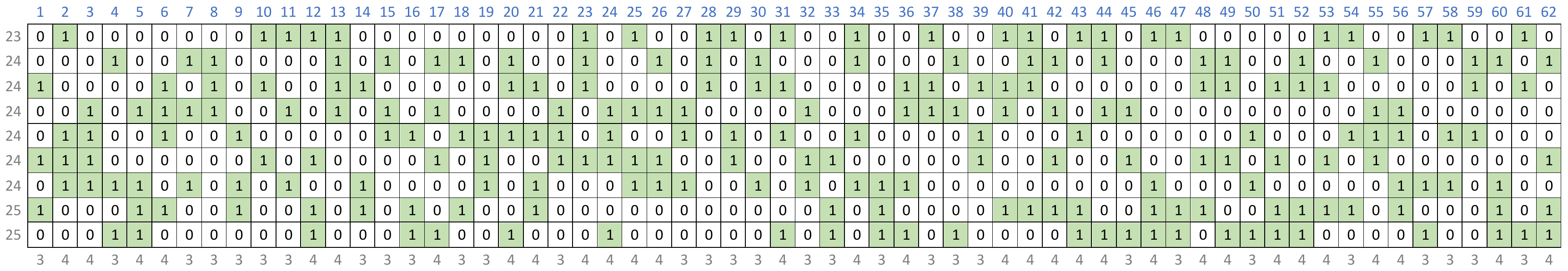}\vspace{-0.45cm}
	\begin{center}
	    {\footnotesize (a) $9\times 62$ sensing matrix}\vspace{0.45cm}
	\end{center}
	\includegraphics[width=6.5in]{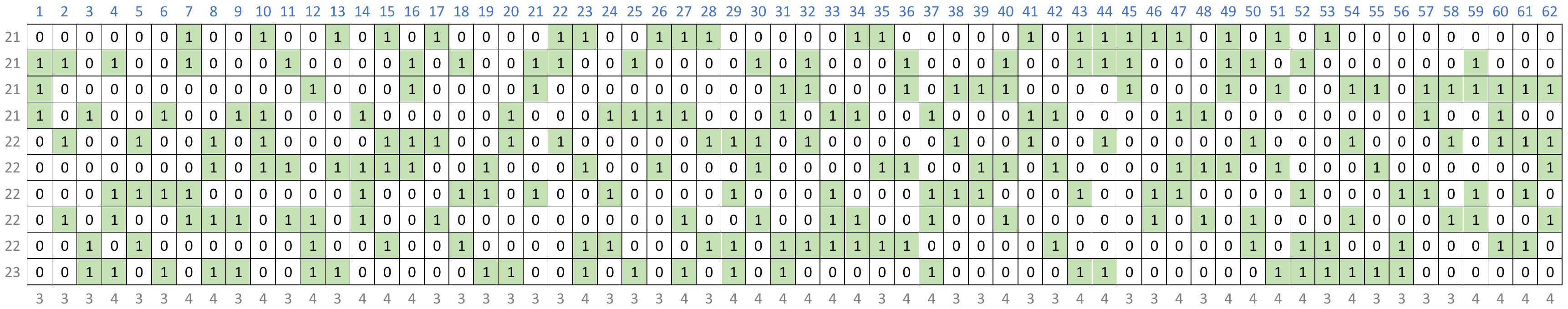}\vspace{-0.45cm}
	\begin{center}
	    {\footnotesize (b) $10\times 62$ sensing matrix}\vspace{0.45cm}
	\end{center}
	\includegraphics[width=6.5in]{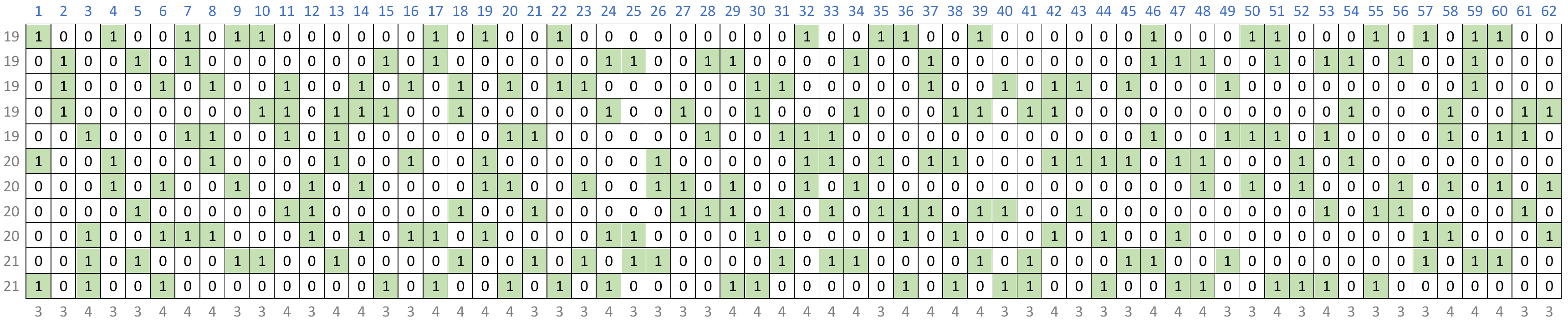}\vspace{-0.45cm}
	\begin{center}
	    {\footnotesize(c) $11\times 62$ sensing matrix}
	\end{center}
	\caption{The sensing matrices with $9,10,11$ rows and $62$ columns used in the simulations for the 2-STAMP scheme. (The weight of each row is shown in front of the row, and the weight of each column is shown below the column.)}\label{fig:MatricesIII}
\end{figure}

For the 2-STAMP scheme, we used the mixing sparsity threshold $\kappa=2$. The number of measurements $m^{(2)}_l$ used in the second stage for each positive pool $l$ such that $\widehat{k}_l\geq 3$ was chosen the same as that in Scheme~I, i.e., $m^{(2)}_l = 7$ for $\widehat{k}_l = 3$, and $m^{(2)}_l = 8$ for $\widehat{k}_l \geq 4$. For each (mixed) super-pool $l$ -- formed by combining two positive pools $l_1$ and $l_2$ -- the number of measurements $m^{(2)}_l$ used in the second stage was chosen as follows (depending on both $\widehat{k}_{l_1}$ and $\widehat{k}_{l_2}$):     
\[m^{(2)}_l = 
\begin{cases}
9, & (\widehat{k}_{l_1},\widehat{k}_{l_2}) = (1,1),\\
10, & (\widehat{k}_{l_1},\widehat{k}_{l_2}) = (2,1),\\
11, & (\widehat{k}_{l_1},\widehat{k}_{l_2}) = (2,2).\\
\end{cases}
\] For each $m^{(2)}_l$, the column and row weight distributions of the $m^{(2)}_l\times 62$ sensing matrix $\mathbf{A}^{(2)}_l$ used in the simulations are described below (see Fig.~\ref{fig:MatricesIII}):  
\begin{itemize}
    \item $m^{(2)}_l=9$: 
    \begin{itemize}
        \item $31,31$ columns of weights $3,4$, respectively;
        \item $1,6,2$ rows of weights $23,24,25$, respectively.
    \end{itemize}
    \item $m^{(2)}_l=10$: 
    \begin{itemize}
        \item $31,31$ columns of weights $3,4$, respectively;
        \item $4,5,1$ rows of weights $21,22,23$, respectively.
    \end{itemize}
    \item $m^{(2)}_l=11$: 
    \begin{itemize}
        \item $31,31$ columns of weights $3,4$, respectively;
        \item $5,4,2$ rows of weights $19,20,21$, respectively.
    \end{itemize}
\end{itemize}

\end{document}